# Enterprise Resource Planning Systems: the Integrated Approach

Sergey V. Zykov
ITERANET Co. Ltd.,
ITERA International Group of Companies,
Moscow, Russia
szykov@itera.ru

**Abstract**

Enterprise resource planning (ERP) systems enjoy an increasingly wide coverage. However, no truly integrate solution has been proposed as yet. ERP classification is given. Recent trends in commercial systems are analyzed on the basis of human resources (HR) management software. An innovative "straight through" design and implementation process of an open, secure, and scalable integrated event-driven enterprise solution is suggested. Implementation results are presented.

## 1. Introduction

Frequent priority changes in versatile corporations development demand fast and flexible adaptation of personnel organizational structure to rapidly changing modern market with stiff competition. Such adaptation should be based on strategic software integration, and is especially urgent for comprehensive ERP systems that involve a collection of technologies allowing to support complex production-and-trade cycles.

During the two recent decades, the data models (DM) and architectures underlying software development process have been changed significantly to support object methodologies and interoperability [9]. Since early 90-s, rather uniform file-server systems of versatile material resource planning (MRP) supporting relational [4] DM have been developed up to well-consolidated ERP systems based on extendable relational DM supporting object-relational and object-oriented DBMS. Attempts of enterprise application integration (EAI) have also been undertaken [2,5,6,10,12].

The approach suggested in this paper for ERP integration has been practically approved while prototyping and implementing a full-scale HR information system (HRIS).

Main objectives of the paper are the following ones:

- ERP design methods classification and analysis;
- integrated data and metadata model development;
- ERP component integration algorithm development;
- HRIS prototyping and full-scale implementation.

Research methods meeting the problem domain specific features are based on a creative synthesis of fundamental statements of lambda calculus [18], category [8] and semantic network [16] theories.

The data model is introduced provides integrated problem-oriented event-driven data and metadata dynamics and statics management of heterogeneous weak-structured problem domains in a more adequate way than previously known ones. The integration algorithm suggested allows to generate system architecture solution for open, distributed, interoperable environment supporting front-end versatile data warehouse processing based on dynamic SQL with stored procedures and object-oriented CORBA, UML and business-process reengineering (BPR) technologies.

## 2. ERP History

ERP construction procedures are based on data model, architecture, DBMS and CASE technologies. Let us briefly describe major ERP development approaches.

### 2.1 Legacy Systems

First enterprise-level solutions have been obtained, apparently, in 60-s as custom-made IS for inventory control (IC) based on mainframes. Examples include hierarchical (*IBM IMS*, 1968) and network (*Cullinet Software IDMS*, 1971) systems.

Some of the IC systems included such advanced features as multimedia personal data handling and flexible form and report generation. Security has been restricted by bulky procedures for explicit enabling/disabling rights of every user for each entry form, report or query. Since legacy systems have been based on file-server, they lack flexibility and portability, therefore reengineering and data publishing is a serious problem.

A positive example of such systems is *UniQue* HRIS from *Q Data Dynamique* originally designed for use with AS/400 and later on, in 1993, adapted for PC LANs [21].

However, the legacy IC systems are not based on a data model and therefore they use rather primitive set of standard functions. Neither do they feature front-end programming and development environment, nor can they transparently acquire foreign source data.

## 2.2 From Legacy to High-End: Integration and Flexibility Start Here

### Standards and Models Change: MRP to EAI

In 70-s material resource planning (MRP) systems appeared in order to satisfy emerging demand for enterprise resource management. The MRP systems were based on relational DBMS [4]. The major prototypes included *System R* [11] and *Ingres* [20] which lead to commercial *IBM DB2* and *CA Ingres*. In general, MRP systems support dynamic SQL and provide more flexible resource control than IC systems. MRP solutions are aimed at corporations employing less than 10,000 people.

In early 90-s further development of data models (P.Chen's ERM [3]), data manipulation languages (SQL2), industrial DBMS (*Oracle, Informix, Sybase*) [15] and integrated CASE tools has resulted in advent of client/server ERP systems that provided comprehensive resource management for large corporations with more than 10,000 employees. Later on, general-purpose system integration software solutions based on challenging concepts and tools have been obtained by D.Calvanese [2], D.Florescu, A.Levi [6] (ODBC/JDBC data integration), D.Linticum [12], H.Davis [5] (COM/CORBA application integration) as well as Y.Kambayashi [10] (Java, ActiveX, MOM and RPC interface integration).

The most successful example of state-of-the-art commercial ERP is *Oracle Applications*.

### Oracle Applications: an Integrated ERP Solution

The *Oracle Applications* group includes the following integrated modules: *Human Resources, General Ledger, Accounts Payable, Accounts Receivable, Fixed Assets, Manufacturing, Project Management,* and *Purchasing*.

*Oracle Applications* services are based on highly scalable and reliable relational database *Oracle Universal Server*. The ERP benefits compact storage, effective retrieval of multimedia data, advanced form generator and report writer, object-oriented visual interface script language, SQL-based procedure-oriented query language, cross-platform support, WWW-ready applications development.

Multi-platform client/server support is provided for most of leading operating systems including *MS Windows NT*, *Sun Solaris*, *IBM AIX* and other *UNIX* dialects etc.

Developed and deployed database-oriented applications become Web-enabled through *Oracle Web Server*.

*Oracle Designer/2000* CASE and RAD tool allows to enhance and optimize business processes through visual interface and SQL-based *PL\SQL* language as a basis and a visual object-oriented script language at upper level.

There are certain points of integration between *Oracle Assets* and *Oracle Human Resources* that allow using personal data from for depreciation and tax calculations. However, *Oracle Applications* group products are integrated loosely enough and much is still desired to build a real enterprise level solution out of them.

## 3. Related Works

Papers [1,8,4,9-11,16-19] provide rigorous mathematics foundation and solid theoretical research background for database structure notations. Relational DBMS and weak-structured document solutions are cross-examined.

Object hierarchy and semantic network as a basic approach to handling object storage and database structure manipulation is described in [6,7].

Lattice of flow diagrams, used for data flow modeling, is discussed in [17].

Through rigorous mathematics background papers [22,23] provide an overview of object-oriented systems development and a number of practical solutions.

Enterprise groupware-based solution is outlined in [24] and given a wider coverage in [25-28]. ERP overview is based on user and system manuals [21] directly from vendors. World-recognized independent expert opinion [15] is also considered. Current ERP market status is acquired from WWW sources [13,14].

## 4. Architecture and Interface Requirements

According to problem domain research results, vital issues of problem-oriented ERP construction for integrated corporate resource management have been formulated. In accordance with problems detected, fundamental requirements for versatile enterprise-level software design and implementation have been classified.

Specific features of the problem domain require support for dynamic multilevel personnel restructuring process with multi-alternative assignment-based complex estimation of enterprise activity. Interface requirements set should allow dynamic variation of mandatory input fields, flexible access rights differentiation and constant data integrity support.

In architecture respect, the system should provide interoperability, expandability, and flexible adaptation to problem domain changes as well as data and metadata correction possibility, e.g. through rollback.

## 5. The Integrated Data and Metadata Model

### 5.1 The Data Object Model

Mathematical formalisms existing for problem domains are not fully adequate to dynamics and statics semantic peculiarities. Besides, modern methods of CASE-and-RAD design and implementation of integrated enterprise applications do not result in solutions of a wide application spectrum; the corresponding commercial ERP do not provide a significant degree of complex heterogeneous problem domains data usage.

According to results of research on enterprise personnel management problem domain specific features, a computational data model (DM) based on object calculus has been built. The model is a theoretical method synthesis of finite sequences, categories and semantic networks.

Date objects (DO) of the DM introduced can be represented as follows:

DO = < concept, individual, state >,

where a concept is understood as a collection of functions with the same definition area and the same value range. An individual implies an essence selected by a problem domain expert, who indicates the identifying properties. State changes simulate dynamics of problem domain individuals.

Compared to research results known as yet, the DM suggested enjoys significant advantages of more adequate dynamics and statics mapping of heterogeneous problem domains, as well as support problem-oriented integrated data management. In architecture and interface aspects the DM provides straightforward iterated design of open, distributed, interoperable HRIS based on UML and BPR methodologies. As far as implementation is concerned, information processing from various repository types of heterogeneous enterprise problem domains is supported providing front-end data access based on event-driven procedures and dynamic SQL technologies.

The computational model suggested is based on the two-level conceptualization scheme [22], i.e. process of establishing relationship between concepts of problem domain.

Individuals h, according to the types T assigned, are united in assignment-depending collections, thus making variable domains of sort

$H_T(I) = \{h \mid h : I \rightarrow T\}$,

that simulate problem domain dynamics.

When fixing data model individuals, uniqueness of individualization of data object d from problem domain D by means of the formula $\Phi$ is required:

$\| Ix\ \Phi\ (x) \| i = d \Leftrightarrow \{d\} = \{d \in D \mid \|\Phi(d)\| i = 1\}$.

### 5.2 The Metadata Object Model

Compression principle for the computational data object model introduced

$C = Iy: [D]\ x : D(y(x) \leftrightarrow \Phi\ ) = \{x : D \mid \Phi\}$

allows to apply the model to concepts, individuals and states separately, as well as to data objects as a whole.

The computational metadata model expands traditional ER-model [4] by a principle of compression:

$x^{j+1}\ Iz^{j+1}: [\ldots[D]\ldots]\ \forall x^j: [\ldots[D]\ldots]\ (z^{j+1}(x^j)\ \Phi^j)$, where

$z^{j+1}, x^{j+1}$ – metadata predicate characters in relation to level j,

$x^j$ - individual of level j,

$\Phi^j$ - data object definition language construction of level j.

The suggested comprehensive model of objects of the data, metadata and states is characterized by scalability, aggregation, metadata encapsulation, hierarchy structure and visualization.

Expandability, adequacy, neutrality and semantic correctness of the formalism introduced provide problem-oriented software design with adequacy maintenance at all stages of implementation.

Semantics of computational model of objects of the data, metadata and states is adequately and uniformly formalized by means of typed λ-calculus, combinatory logic, and semantic network-based scenario description.

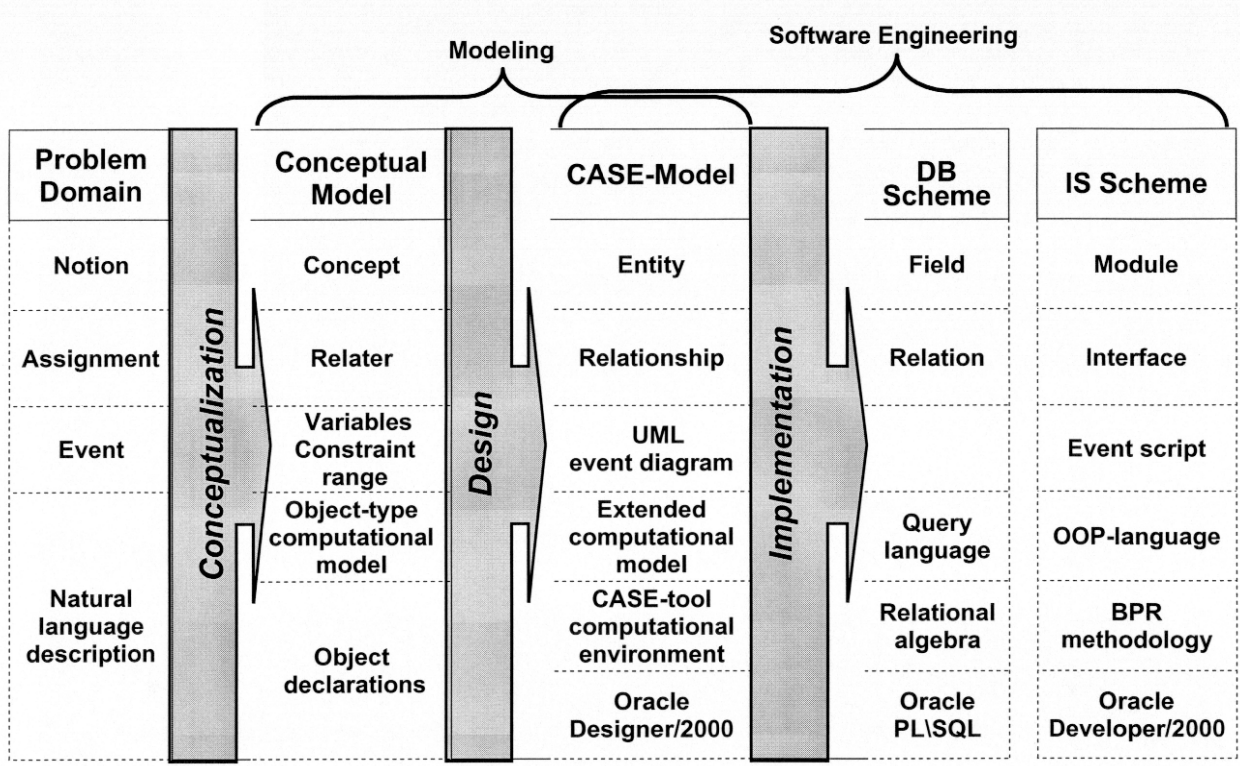

Figure 1 Generalized implementation scheme for enterprise information systems

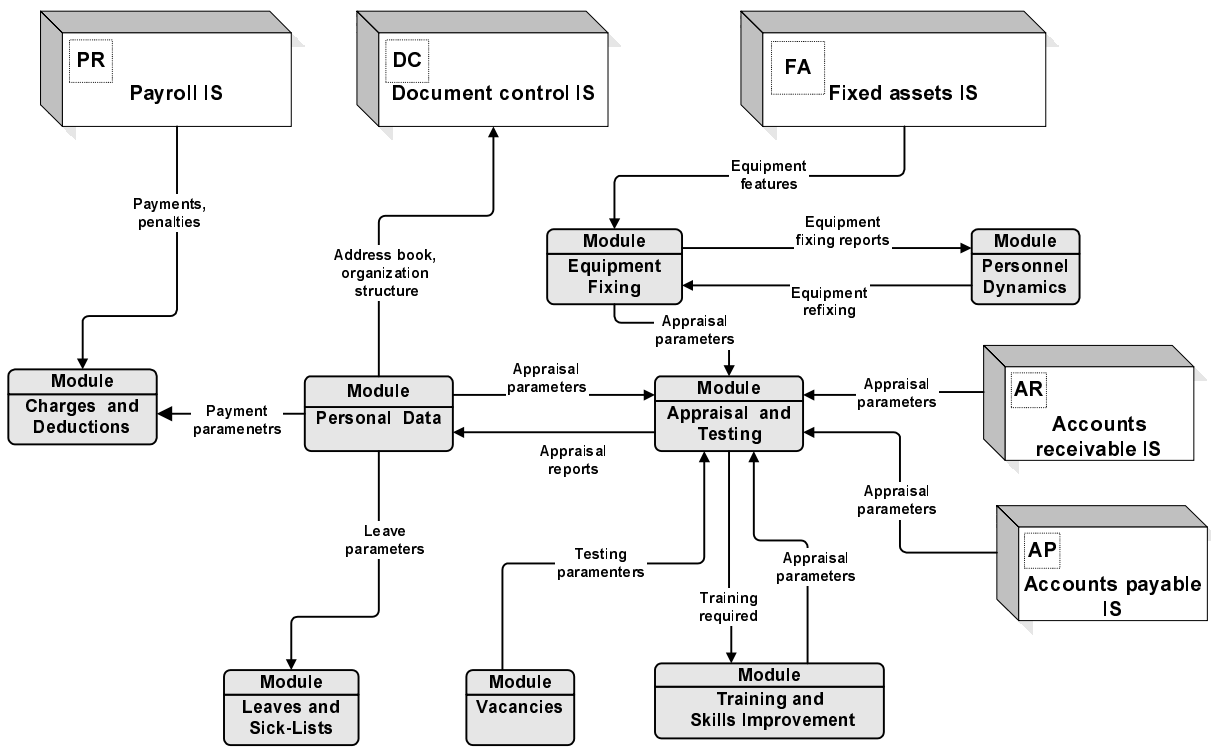

Figure 2 Implemented enterprise HRIS structure

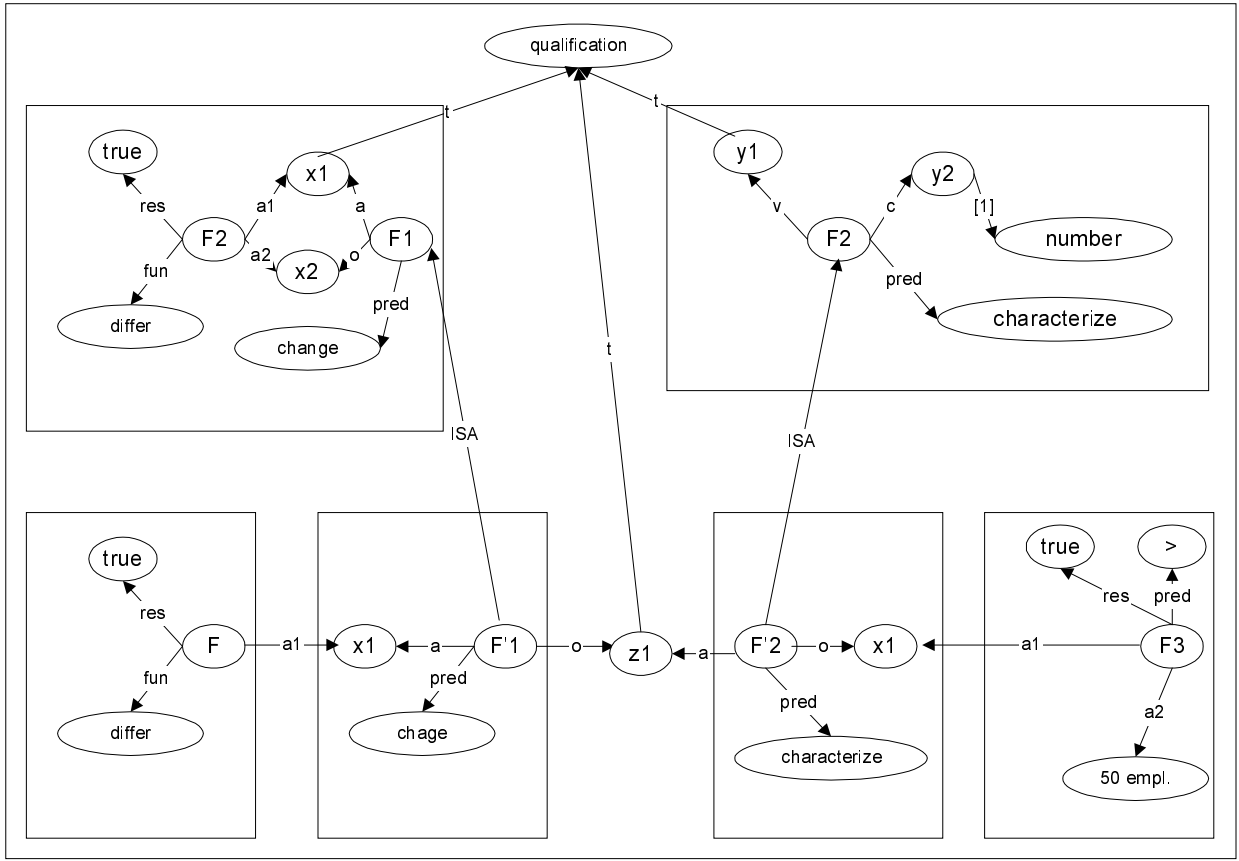

**Figure 3 Situation business-model fragment in the form of the semantic network**

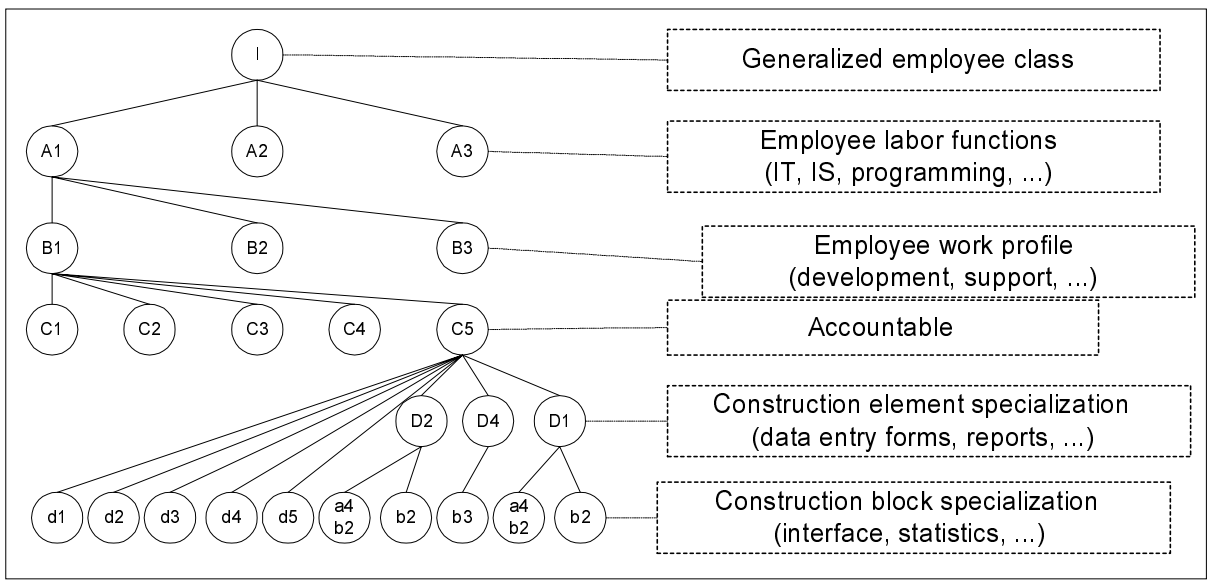

**Figure 4 Levels of generalized personnel class hierarchy**

## 5.3 The Integrated Model Application to ERP

### Enhancing HR Component: Personnel Appraisal Model

Let us introduce an appraisal modeling formalism meeting the general requirements for specific features detected during ERP problem domain analysis. In the below example most essential employee evaluation parameters:

- hierarchy-based corporation organizational structure;
- employee working functions;
- vacant positions currently available;
- enrolled employee amount
- are combined.

The last three parameters can be detailed from the corporation level down to its smallest structural unit, while the first parameter is a global one.

Let us assume that A and B are sets.

Let $B^A$ stand for the mapping from A to B:

$B^A = \{f | f: A \rightarrow B\}$.

Let us match $B^A$ with the mapping

$||\circ|| = \{f \ | \ f \ B^A \times A \rightarrow B\}$,

i.e. with $||\circ||$ evaluation function.

Thus, the following equation holds true

$||\circ|| = (<f, x>) = f(x)$, or, in the other form,

$||<f, x>|| = f(x)$.

Now, let us build the semantics network language model. Let us consider an ordered pair of DO of the form

L=<R,C>, where

$R=\{R_1, R_2, \ldots\}$ stands for predicate dyadic predicate symbols set and

$C=\{C_1, C_2, \ldots\}$ stands for constants set.

Therewith, the atomic formulae of the model suggested correspond to simple frames, and terms denote problem domain individuals.

Let us construct frame evaluation procedure using the introduced evaluation function $||\circ||$. Let us also relate the semantic network-based DM with databases in general, and with relational databases in particular.

Let us consider the following natural language situation of HR management: "An General_Director employee is changing his position to Department_Director, managing a department of more than 50 employees" (see fig.3).

Now let us consider an example of employee activity evaluation based on the formal DM (see fig.4).

Let the F functional denote the most general class of employees. Let the assignment s={development, support} accounts for corporation personnel labor functions.

Let F(s) stand for the set of employees, for whom the labor functions are restricted to development and support.

Let the assignment p={information technologies, programming, information systems} accounts for department-level organization unit.

Let us name as F(s)(p) the set of employees with development and support labor functions for whom an organization unit is assigned i.e., currently enrolled ones. For the sake of simplicity and without loss of generality, let us consider that in general the employee set (i.e., corporate personnel), referred above as functional F is dependent upon vacancy number (v), number of staff members (e), their labor functions and organization units (i.e., on organization structure):

$F=F((v), (e), \ldots)$.

In this case, the formula $||F=F((v), (e), \ldots)||$ indicates a formal procedure that evaluates parameterized functional, the expression $||F=F((v), (e), \ldots)(s)||$ evaluates employees with given labor functions (s), and the formal "procedure" $||F=F((v), (e), \ldots)(s)(p)||$ evaluates employees with given labor functions (s) from organization units (p).

The introduced functional F can be considered a foundation for a computational formalism of parameterized procedure of comprehensive appraisal of a certain corporation organization unit level (from companies and departments to employees).

Let us demonstrate that two-level conceptualization scheme is sufficient for the model adequacy.

Let us introduce the following denotations:

$||r|| = \{r_{l.f.}, r_{o.u.}\}$ – specific costs;

$||z|| = \{z_{l.f.}, z_{o.u.}\}$ – segmentation degree (i.e., possibility of dividing the personnel into stable and independent work groups)

$||q_i|| = q_i$ – overheads;

$||l_i|| = l_i$ – work (contract, project, order, target goal) stage duration;

$\|n_i\| = n_i$ – number of work stages.

Evaluated values are generalized, i.e., there is no uniqueness of value choice for specific costs and segmentation degree.

Generalization level decrease is achieved by assignment point s consideration:

$$\|z\|(\|s\|) = \begin{cases} \|z\|(\text{devel.}) = z_{\text{devel.}}, \\ \|z\|(\text{support}) = z_{\text{support}}; \end{cases}$$

$$\|r\|(\|s\|) = \begin{cases} \|r\|(\text{devel.}) = r_{\text{devel.}}, \\ \|r\|(\text{support}) = r_{\text{support}}. \end{cases}$$

Moreover, further generalization level decrease by considering the second assignment point p does not result in success:

$$\|z\|(\|s\|)(\|p\|) = \|z\|(\|s\|);$$
$$\|r\|(\|s\|)(\|p\|) = \|r\|(\|s\|).$$

The result obtained can be explained by the fact that the evaluation procedure does explicitly include organization structure position.

However, it is obvious that overheads $q_i$ are dependent both on labor functions and on organization structure position, i.e. we should let $\|q_i\| = \{q_{i\text{ devel.}}, q_{i\text{ support.}}\}$.

The equality $\|q_i\| = q_i$ implies that

$q_{i\text{ devel.}} = q_{i\text{ support}} = q_i$.

**Generalized Component Integration Algorithm**

According to the ERP design and implementation scheme (see fig.1), a generalized algorithm of new component integration into existing ERP structure is suggested.

The algorithm is based on the semantically preferred data objects analysis and provides consistency and integrity of extendable data object models as well as possibility of iterated information system design through by business model reengineering. The algorithm unifies object-based heterogeneous management information system integration process.

An important feature of the generalized component-integration algorithm is its semantic orientation. In terms of human resources information management system it implies organizational structure dependency.

The author has performed research of corporation organizational structure semantics earlier. Research results are presented in [25-28], where they are discussed in a more detailed way.

## 6. The Integrated ERP Implementation

### 6.1 Customizing the Implementation Scheme

During design process, ERP specification is transformed from problem domain concepts to data model essences and further through CASE-tools to DBMS scheme with *PL\SQL* as data object manipulation language to target ERP description with appropriate architectural and interface components (see fig.1). In accordance with problem domain specific features analysis results, computational data model and generalized scheme of ERP development have been adapted to satisfy the required personnel management conditions. The problem-domain oriented ERP design scheme includes five stages:

- corporation board of directors formulates objectives, measures and plans on restructuring which are mapped in formal business rules of ERP computational model (see fig.5);

- experts in personnel and information create the specified structural and functional conceptual corporation business model in a form of object map;

- system analysts make OLAP-research of corporate business model variants for various development scenarios;

- database and ERP developers formalize business logic of the architecture and interfaces using object-oriented script language, language, which is assembled in UML-data model by means of CASE-synthesis methodology;

- database, local area network and security managers implement and support target ERP and DB schemes.

### 6.2 Problem-Oriented Interface and Event-Driven Architecture

According to detailed ERP design sequence, a generalized heterogeneous repository processing scheme is introduced that allows users to interact with distributed database in a certain state depending on dynamically activated (i.e., assigned) scripts. Thus, the scripts in a form of database connection profiles and stored object-oriented program language procedures are initiated depending on user-triggered events. Scripts provide transparent and intellectual client/server front-end user-to database connection. Dynamically varied database access profiles provide high fault tolerance and data security both for ordinary and privileged system users in the heterogeneous environment. The profiles are implemented using CORBA technology as an intellectual media between end user and heterogeneous data warehouses.

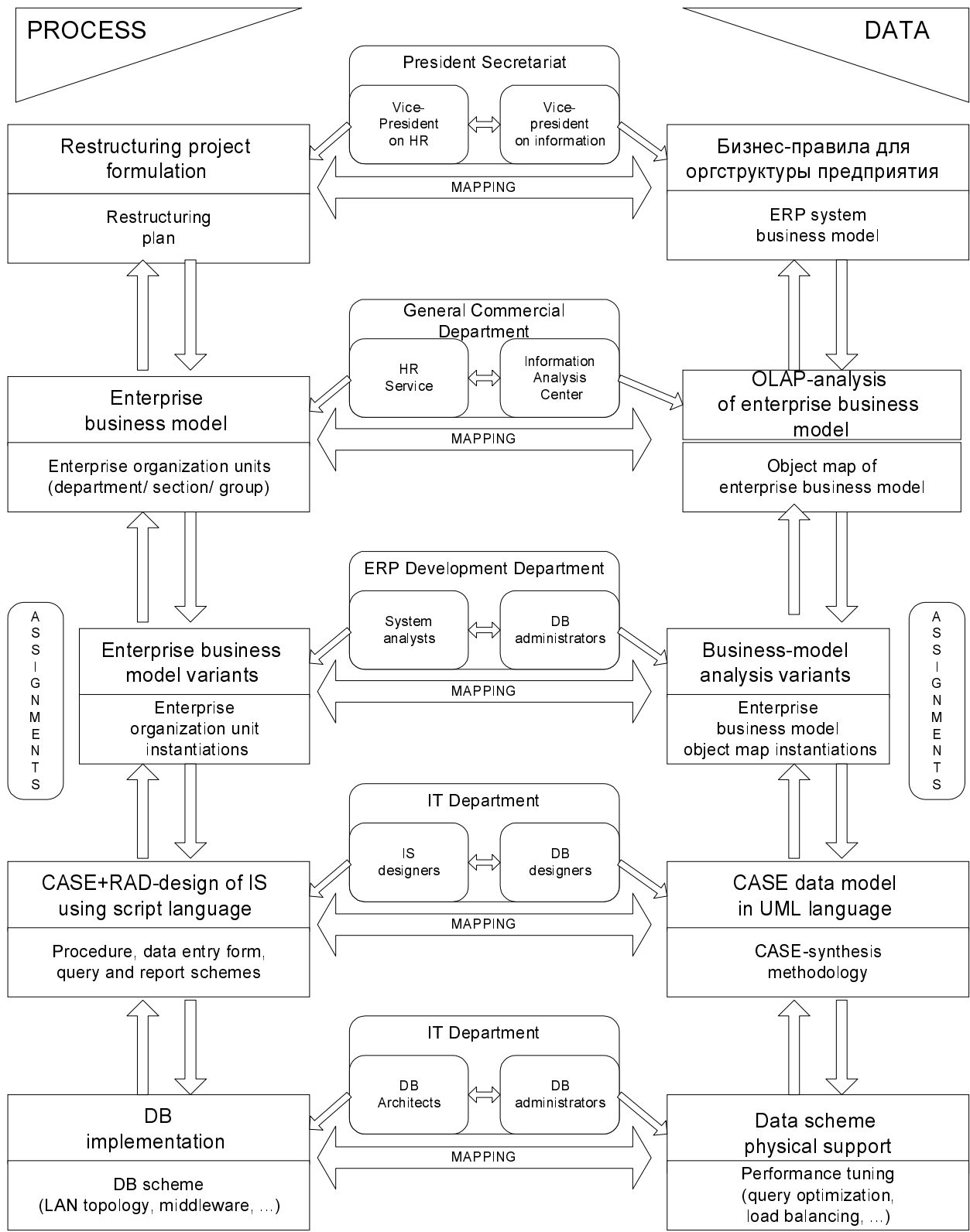

**Figure 5 ERP system design and implementation levels**

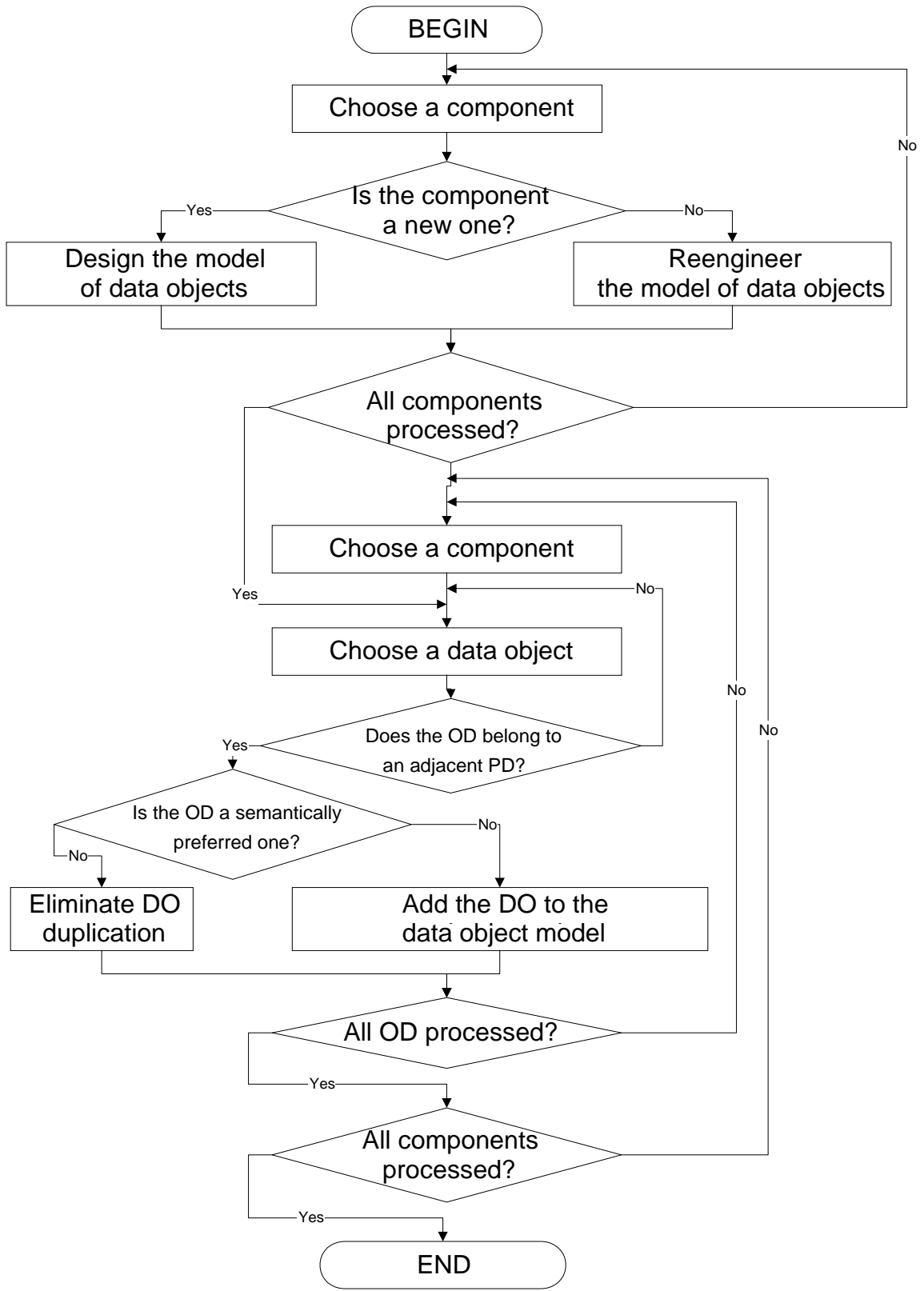

**Figure 6 ERP integration algorithm**

Depending on semantic-oriented corporation organizational structure, which defines ERP user position in the corporation hierarchy, a certain database connection and access level profile is assigned dynamically. The profile is valid only until the end of a data exchange session. According to the corporation hierarchy, user gets access to data under one of the basic scenario profiles ranging from corporation president to a department employee. Access is granted not only to data, but also to metadata (i.e., data object dimensions, integrity constraints, access rights, appraisal parameters, corporation structure etc.).

Administrative users have extended access to metadata. Thus, under the model, introduced data and metadata objects are manipulated uniformly. This makes system interface a problem-oriented, straightforward and uniform one and significantly increases system performance and user-friendliness.

## 6.3 Implementation Description

The introduced design methodology has been practically approved during ERP HR component implementation at *ITERA* International Group of Companies. The enterprise personnel management software consists of eight components (see fig.2). *Personal Data* component is intended for storage and processing of employee biography data. The subsystem, connected to the above one, named *Personnel Dynamics* allows to control dynamics of employees enrollment, transfer, dismissal and re-enrollment events. The adjacent software module *Charges and Deductions* provides registration of salaries, bonuses and other kinds of payments, as well as material penalties. The central *Appraisal and testing* component supports comprehensive employee labor activity estimation based on individual, psychological, professional and other kinds of tests, as well as on adjacent modules and third-party software data. *Vacancies* component supervises personnel selection according to given criteria. *Leaves and Sick-Lists* subsystem accounts employees working hours and supports multi-type leaves. *Training and Skills Improvement* component implements training policy judging by appraisal results and tracks training service payments. Finally, *Equipment Fixing* subsystem provides registration for accountable persons and major corporation resources used by them.

All of the HR component modules are captured by unified interface and integrated into ERP environment of the *Oracle Applications* financial and commodity management systems and *Oracle InterOffice* document management system.

From the system architecture viewpoint the integrated HR component provides certain level of data input, correction, analysis and output (from president down to chief of a department) depending on front-end position (i.e., assignment) in personnel hierarchy. Problem-oriented form designer, report generator, on-line documentation and administration tools are used as interactive interface facilities. The ERP database supports the integrated storage for data (for on-line user access) and for metadata (data object dimensions, integrity constraints and other business process parameters).

During the ERP design process problem domain data model specification (represented as semantic network fragments) has been transformed into use-case UML diagrams, then, by means of *Oracle Developer/2000* integrated CASE-tool - into ER-diagrams and, finally, into the attributes of target ERP and databases.

On the basis of the information model developed, architecture-and-interface solution for integrated personnel management software has been designed; details of database processing for various system user and administrator classes have been considered.

Software implementation has been divided into two stages: 1) fast prototype created with an SQL-based query language, supporting triggers and stored procedure mechanisms using *PowerScript* script object-oriented language and 2) full scale and capacity software implementation based on the *Oracle* integrated information system development tools platform.

To prove adequacy of the computational data and metadata model developed and component integration algorithm suggested, a fast software prototype has been designed on the basis of generalized architecture scheme and supporting interfaces.

*Sybase S-Designor/PowerBuilder* has been chosen as CASE- and RAD-toolkit for implementation environment as a result of carried out comparative analysis.

According to prototype approbation results full-scale object-oriented software has been implemented and subsequently adapted for personnel management application development.

To provide required levels of industrial scalability and fault tolerance, judging by the results of CASE-and-RAD tools multi-criteria comparative analysis *Oracle Developer/2000* toolkit has been chosen as an integrated solution supporting methodologies of universal modeling (UML) and business processes reengineering (BPR) methodologies.

Target ERP implemented consists of eight components using a set of Oracle tools. All of the components are implemented according to technical specifications designed by the author personally, and amount to more than 1000 source text pages. 150 high complexity bilingual screen forms and reports, as well as about 30-page source text size *Equipment Assignment* component have been also created by the author. According to specification requirements developed by the author together with *ITERA* International Group of Companies personnel service the software implemented had been significantly improved. In particular, procedures of accounting salaries and vacation bonus have been coded.

The full-scale implementation is based on the hardware platform of an IBM RS/6000 two-server high availability cluster running under AIX operating system.

The information system has been implemented in a large international corporation and has passed a three-year experimental check.

As a result of software implementation designed on the basis of the model introduced, implementation terms and cost and cost compared to existing commercial software of the kind are considerably reduced while the functional set is extended.

In the opinion of users, the software implemented features high degrees of openness, expandability, flexibility, reliability, ergonomics and ease of mastering.

Thanks to problem-oriented interface, primary data entry speed exceeds that of commercial software of the kind by the average of 20% and amounts to about 150 seconds per employee data entry. Access levels differentiation allows to considerably result risk of information distortion or loss.

## 7. Results

Computational data model has been introduced providing integrated manipulation of data and metadata objects especially under the conditions of rapidly changing heterogeneous problem domains. The model is an alloy of methods of finite sequences, category theory and semantic networks.

An original generalized scheme of "straightforward" enterprise ERP design and implementation has been proposed on the basis of formal data and metadata model. The scheme includes a for new components integration into an ERP that provides adequacy, consistency and data integrity.

According to the above mentioned scheme and algorithm, a generalized ERP interface has been designed based on an open and extendable architecture.

To solve a complex applied task of enterprise resource management, fast event-driven prototype software has been developed on the basis of generalized interface and architecture of structural and logical UML scheme.

Using the prototype approbation results, a full-scale object-oriented ERP application has been designed and applied for a versatile enterprise-level implementation.

The full-scale enterprise-level software has been customized for corporate resource management and implemented at a corporation with more than 1000 employees.

## 8. Conclusion

Results for integrated ERP solution implemented have proved significant decrease in terms and costs of implementation as well as growth of portability, expandability, scalability and ergonomics levels in comparison with existing commercial software of the kind. Iterated multilevel software design is based on formal model synthesizing object-oriented methods of data (data objects) and knowledge (metadata objects) management. Industrial implementation of the ERP HR component has been carried using integrated CASE- and RAD-toolkits. Practical implementation experience has proved importance, urgency, originality and efficiency of the approach as a whole as well as of its separate stages and solutions.

Theoretical and practical statements outlined in the paper have been approved by enterprise-level ERP software successful implementation at *ITERA* International Group of Companies.